\documentclass[onecolumn,9pt]{IEEEtran}

\usepackage{url}
\usepackage{./bs_macros}
\usepackage{caption}
\usepackage{float}
\usepackage[labelformat=simple]{subcaption}

\usepackage{balance}

\graphicspath{{./}{./Figures/}}

\theoremstyle{definition}

\makeatletter
\patchcmd{\@maketitle}
{\addvspace{0.5\baselineskip}\egroup}
{\addvspace{-1.3\baselineskip}\egroup}
{}
{}
\makeatother

\begin{document}

\title{\large  \textbf{Graph Theory and Metro Traffic Modelling}}

%
%
%
%
%

\author{
	
Bruno Scalzo Dees, Anthony G. Constantinides, Danilo P. Mandic
	
\thanks{The authors are with the Department of Electrical Engineering, Imperial College London, UK, e-mail: \{bs1912, a.constantinides, d.mandic\}@imperial.ac.uk.}


}


\maketitle

%
%
%
%

\IEEEpeerreviewmaketitle

\noindent With the rapid development of many economies, an increasing proportion of the world's population is moving to cities, and as such urban traffic congestion is becoming a serious issue. For example, underground traffic networks routinely undergo general maintenance, frequently exhibit signal failures and train derailments, and may even occasionally experience emergency measures because of various accidents. These events ultimately require the closure of at least one station which may severely impact the traffic service across the entire network. The economic costs of these transport delays to central London business is estimated to be $\pounds 1.2$ billion per year. Hence, appropriate and physically meaningful tools to understand, quantify, and plan for the resilience of these traffic networks to disruptions are much needed.

In 1926 a map-maker named Fred Stingemore set out to produce a map of the London underground by regularising the spacing between stations, and allowing himself some artistic licence with the routes of the various lines. Stingemore's work was further abstracted by Harry Beck in 1933 to the Graph form we have today. In this article we demonstrate how graph theory can be used to identify those stations in the London underground network which have the greatest influence on the functionality of the traffic, and proceed, in an innovative way, to assess the impact of a station closure on service levels across the city. Such underground network vulnerability analysis offers the opportunity to analyse, optimize and enhance the connectivity of the London underground network in a mathematically tractable and physically meaningful manner.

\section{Traffic centrality as a graph-theoretic measure}

\begin{figure}[H]
	\centering
	\includegraphics[width=0.5\textwidth, trim={0.3cm 0.4cm 0.3cm 0.4cm}, clip]{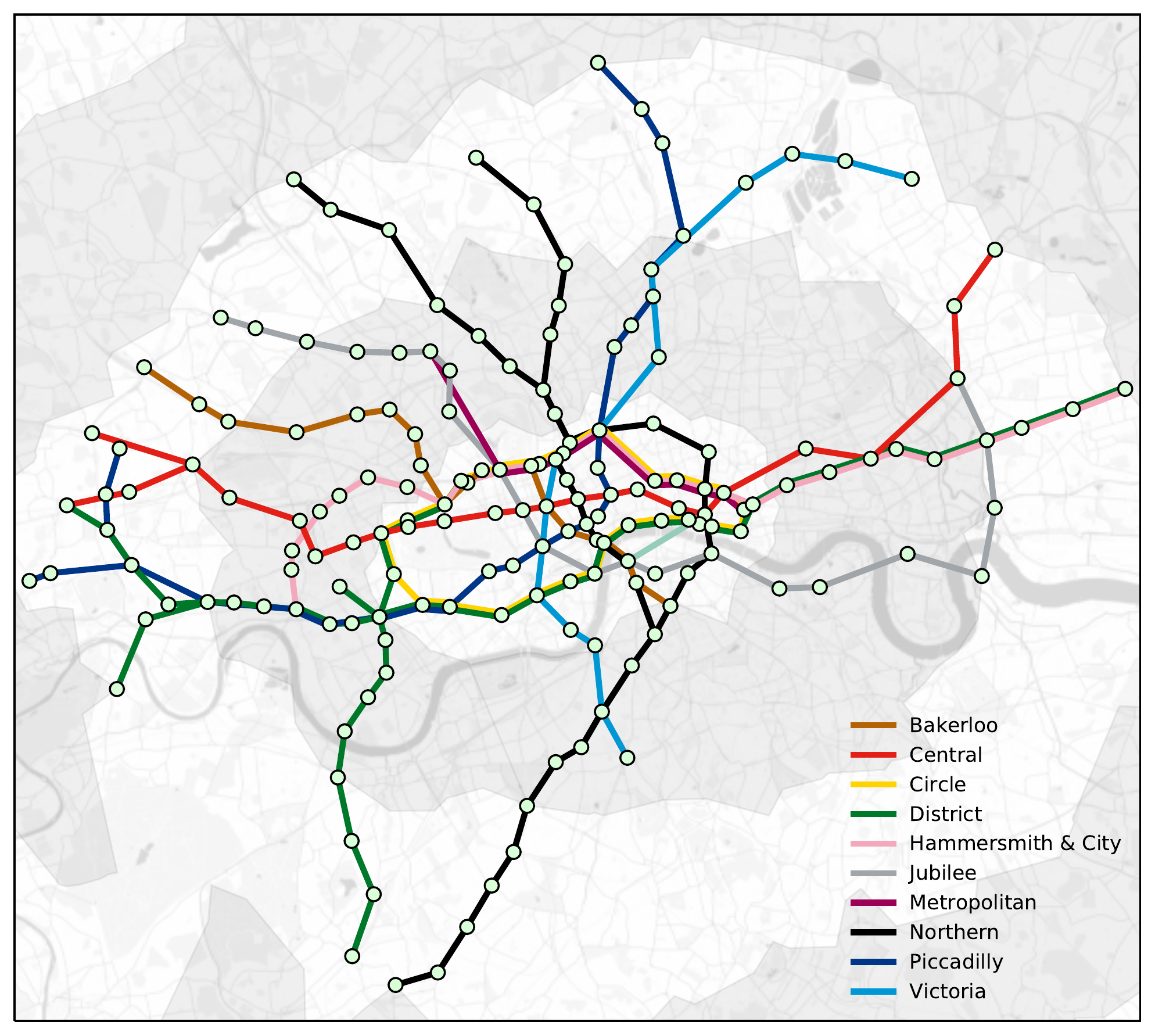}
	\caption{\label{fig:underground_graph}Graph model of the London underground network in Zones 1--3.}
\end{figure}

The underground network can be modelled as an undirected $N$-vertex graph, denoted by $\mathcal{G} = \{ \mathcal{V}, \mathcal{E} \}$, with $\mathcal{V}$ being the set of $N$ vertices (stations) and $\mathcal{E}$ the set of edges (underground lines) connecting the vertices (stations). The connectivity of the network is designated by the (undirected) adjacency matrix, $\mathbf{A} \in \mathbb{R}^{N \times N}$. Figure \ref{fig:underground_graph} shows the proposed graph model of the London underground network, with each vertex representing a station, and each edge representing the underground line connecting two adjacent stations. Notice that standard data analytics domains are ill-equipped to deal with this class of problems.

We employ the following metrics to characterize the topology of the network and model its vulnerability:
\begin{itemize}
	\item  \textit{Betweenness centrality}, which reflects the extent to which a given vertex lies in between pairs or groups of other vertices of the graph, and is given by
	\begin{align}
		B_{n} = \sum_{k,m \in \mathcal{V}} \frac{\sigma(k,m|n)}{\sigma(k,m)}
	\end{align}
	where $\sigma(k,m)$ denotes the number of shortest paths between vertices $k$ and $m$, and $\sigma(k,m|n)$ the number of those paths passing through vertex $n$ \cite{Freeman1977}. In terms of the actual metro traffic, this can also be interpreted as the extent to which a vertex is an intermediate in the communication over the network. Figure \ref{fig:betweenness} shows that, as expected, the stations at the centre of the city exhibit the largest betweenness centrality, and would therefore severely impact the communication over the underground network if disconnected.
	
	\item \textit{Closeness vitality}, which represents the change in the sum of distances between all vertex pairs after excluding the $n$-th vertex \cite{Brandes2005}. Figure \ref{fig:vitality} shows that the stations located in the more remote areas of Zones 2--3 exhibit the largest closeness vitality measure. This is because their removal from the network would disconnect the stations located at the boundaries from the rest of the network.
	
	\hspace{-0.8cm}
	\begin{minipage}[t]{0.5\textwidth}
		\begin{figure}[H]
			\centering
			\includegraphics[width=1\textwidth, trim={0.3cm 0.4cm 0.3cm 0.4cm}, clip]{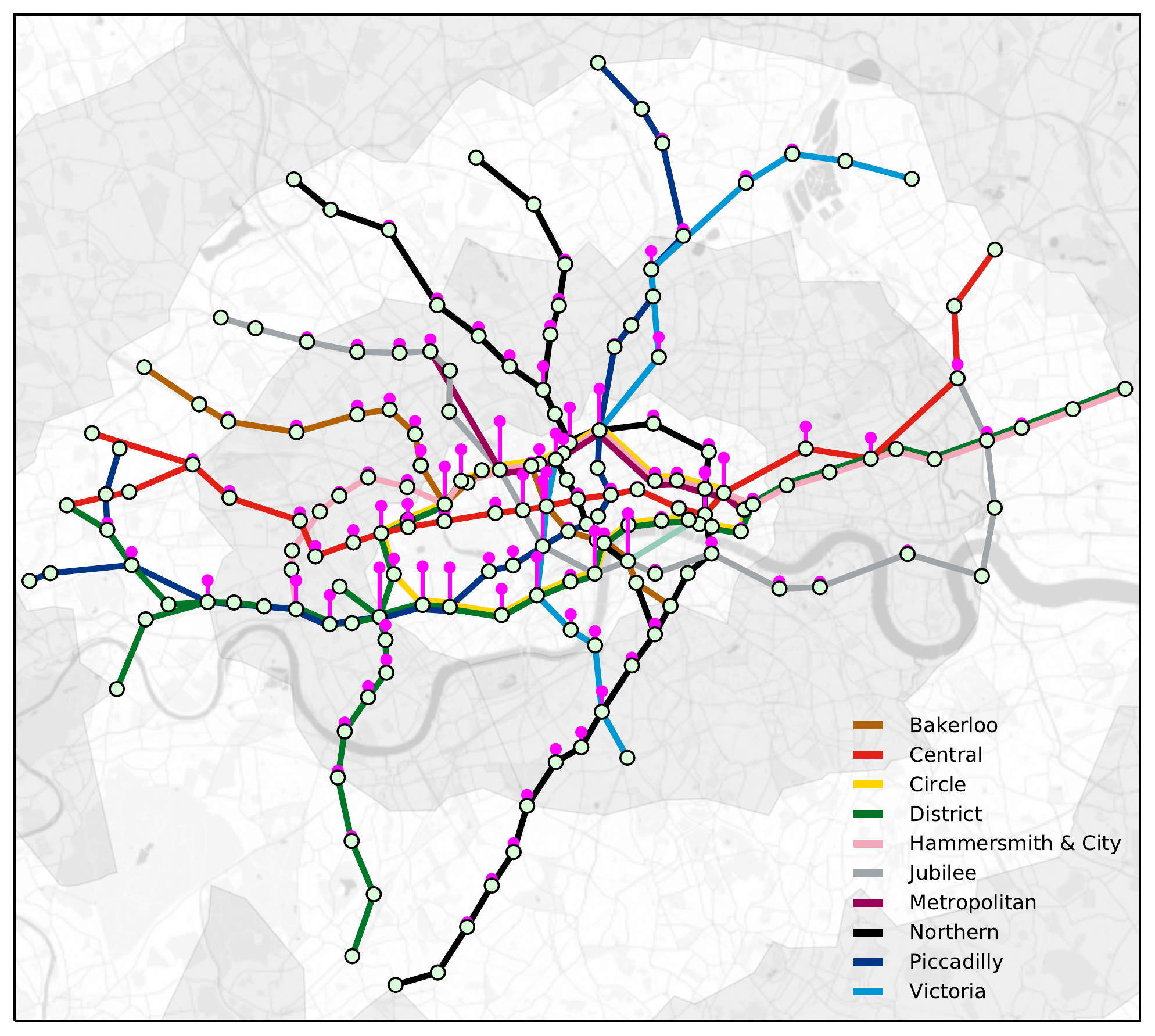}
			\caption{\label{fig:betweenness}Betweenness centrality, designated by magenta-coloured bars, of the London underground network in Zones 1--3. The largest betweenness centrality is observed for the following stations: Green Park, Earl's Court, Baker Street, Waterloo and Westminster.}
		\end{figure}
	\end{minipage}
	\hspace{0.2cm}
	\begin{minipage}[t]{0.5\textwidth}
		\begin{figure}[H]
			\centering
			\includegraphics[width=1\textwidth, trim={0.3cm 0.4cm 0.3cm 0.4cm}, clip]{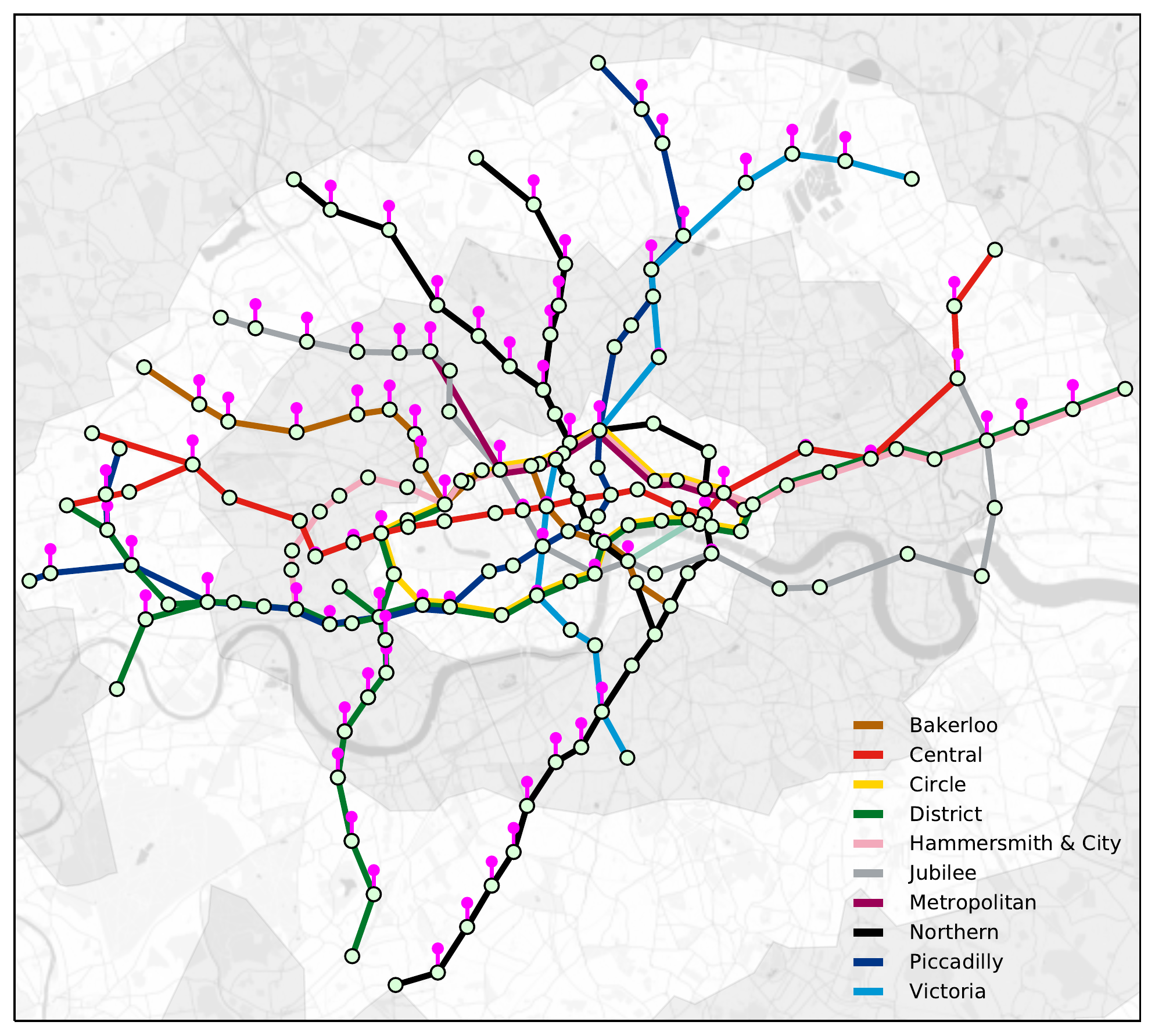}
			\caption{\label{fig:vitality}Closeness vitality, designated in magenta bars, of the London underground network in Zones 1--3.}
		\end{figure}
	\end{minipage}
%
%
	
%
	
\end{itemize}


\section{Modelling commuter population from net passenger flow}

In this section, we employ graph theory to analyse the net passenger flow at all stations of the London underground network. In particular, we demonstrate that it is possible to infer the resident population surrounding each station based on the net passenger flow during the morning rush hour alone.

\begin{figure}[H]
	\centering
	\includegraphics[width=0.3\textwidth, trim={0 3cm 0 4cm}, clip]{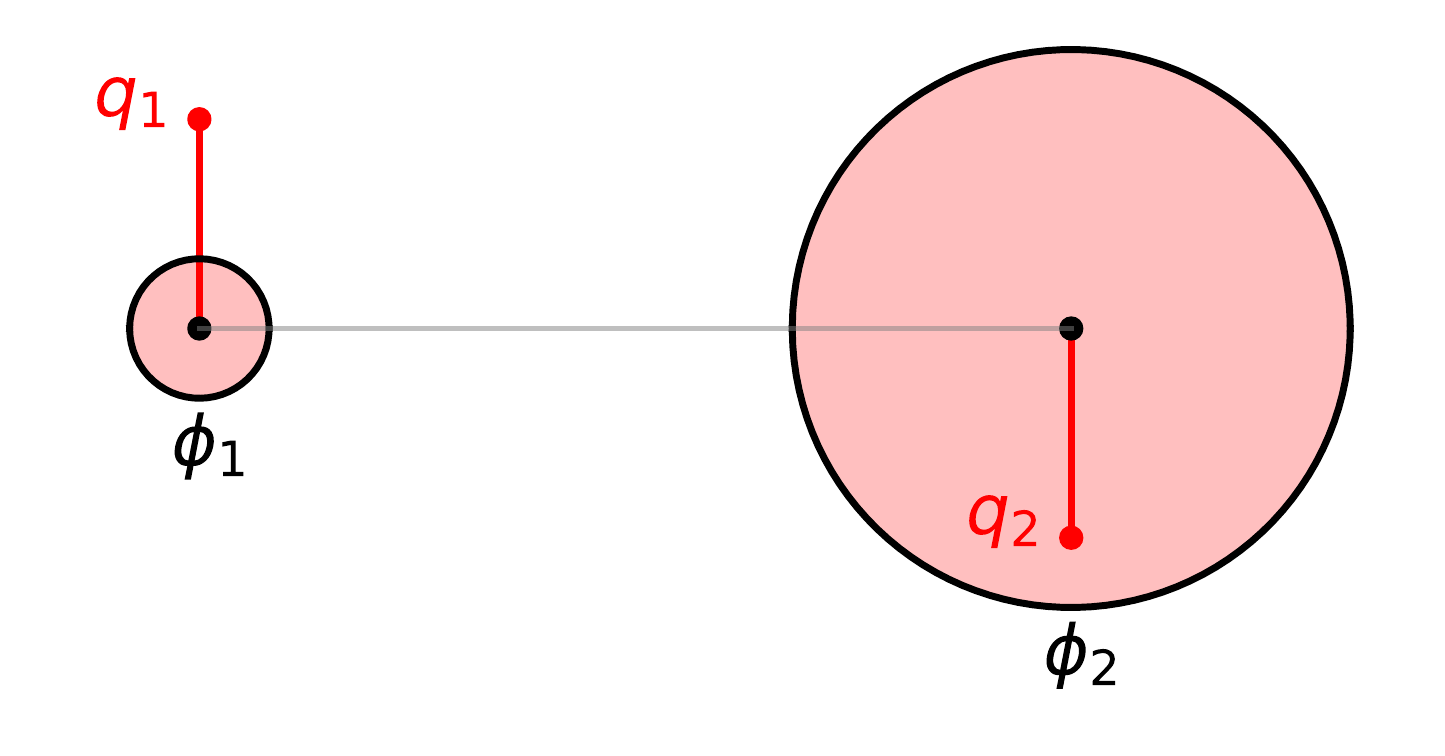}
	\caption{\label{fig:ficks_law}Towards a graph representation of the London underground network. A simplified path graph with two stations surrounded by the respective populations, $\phi_{1}$ and $\phi_{2}$ (proportional to the circle surface area), exhibits the corresponding net fluxes, $q_{1}$ and $q_{2}$. Intuitively, stations surrounded by large populations experience net in-flows of passengers, whereas stations surrounded by low populations experience net out-flows of passengers. Observe that the net flow of passengers across the entire network sums up to zero.}
\end{figure}

To derive the corresponding graph model, we employed the \textit{Fick law of diffusion} which relates the diffusive flux to the concentration of a given vector field, under the assumption of a steady state. This model asserts that the flux flows from regions of high concentration (population) to regions of low concentration (population), with a magnitude that is proportional to the concentration gradient. Mathematically, the Fick law is given by
\begin{align}
	\mathbf{q} = - k \nabla \boldsymbol{\phi}
\end{align}
where 
\begin{itemize}
	\item $\mathbf{q}$ is the flux which measures the amount of substance per unit area per unit time (mol m$^{-2}$ s$^{-1}$);
	\item  $k$ is the coefficient of diffusivity, with its dimension equal to area per unit time (m$^{2}$ s$^{-1}$);
	\item $\boldsymbol{\phi}$ represents the concentration (mol m$^{-3}$).
\end{itemize}
In this way, we can model the passenger flows in the London underground network as a diffusion process, whereby during the morning rush hour the population mainly flows from concentrated residential areas to sparsely populated business districts. Therefore, the variables in our model are:
\begin{itemize}
	\item $\mathbf{q} \in \mathbb{R}^{N}$ is the net passenger flow vector, with the $i$-th entry representing the net passenger flow at the $i$-th station during the morning rush hour, that is
	\begin{align}
		\label{eq:london_underground_flux}
		q_{i} = \text{(passengers exiting station $i$)} - \text{(passengers entering station $i$)}
	\end{align}
	with its dimension equal to ``passengers per station per unit time'';
	\item  $k=1$ is the coefficient of diffusivity, with its dimension equal to ``stations per unit time'';
	\item $\boldsymbol{\phi}  \in \mathbb{R}^{N}$ represents the resident population in the area surrounding the station.
\end{itemize}
This model therefore suggests that, in the morning, the net passenger flow at the $i$-th station, $q_{i}$, is proportional to the population difference between the areas surrounding a station $i$ and the adjacent stations $j$, that is
\begin{align}
	q_{i} & = - k \sum_{j} A_{ij}(\phi_{i} - \phi_{j}) = - k \left( \phi_{i} \sum_{j} A_{ij} - \sum_{j} A_{ij}\phi_{j} \right)  = - k \left( \phi_{i} D_{ii} - \sum_{j} A_{ij} \phi_{j} \right) = - k \sum_{j}\left( \delta_{ij} D_{ii} - A_{ij} \right) \phi_{j} = - k \sum_{j} L_{ij} \phi_{j}
\end{align}
When considering $N$ stations together, we obtain the model in the matrix form
\begin{align}
	\mathbf{q} = -k\mathbf{L}\boldsymbol{\phi} \label{eq:graph_Fick_law}
\end{align}
where $\mathbf{L} = (\mathbf{D}-\mathbf{A}) \in \mathbb{R}^{N \times N}$ is the Laplacian matrix of the graph model. For illustration purposes, Figure \ref{fig:ficks_law} illustrates a signal within this diffusion model on a $2$-vertex path graph obeying the Fick law. 

The data for the average daily net flow of passengers during the morning rush hour at each station in $2016$ was obtained from Transport for London (TFL) \cite{TFL}, and is illustrated as a signal on the underground graph model in Figure \ref{fig:tube_net_flow}. For illustration purposes, Table \ref{table:netflows} shows the daily average net flow of passengers per Zone. As expected, Zone 1 is the only Zone to exhibit a net outflow of passengers, while Zones 2--10 show a net inflow of passengers. In particular, Zone 3 exhibits the largest inflow. In an ideal scenario, the total net outflow across Zones 1--10 should sum up to $0$, however, the residual net outflow is attributed to passengers entering the underground network through other transport services not considered in our model, i.e. rail services. 

Moreover, Table \ref{table:netflows_top5} shows the average net flow of passengers for the top $5$ stations with the greater net inflow and outflow. The stations which the greatest net outflow of passengers are located within financial (Bank, Canary Wharf, Green Park) and commercial (Oxford Circus, Holborn) districts. In contrast, the greatest net inflow of passengers is attributed by railway stations located in residential areas.

To obtain an estimate of the resident population surrounding each station, we can simply set the vector of populations, $\boldsymbol{\phi} \in \mathbb{R}^{N}$, to be the subject of (\ref{eq:graph_Fick_law}), that is
\begin{align}
	\hat{\boldsymbol{\phi}} = -\frac{1}{k} \mathbf{L}^{+}\mathbf{q} \label{eq:population_estimate}
\end{align}
where the symbol $(\cdot)^{+}$ denotes the matrix pseudo-inverse operator. However, notice that the population vector can only be estimated up to a constant, hence the vector $\hat{\boldsymbol{\phi}}$ actually quantifies the \textit{relative} population between stations, whereby the station with the lowest estimated surrounding population takes the value of $0$. The so estimated resident population, based on the morning net passenger flow, is displayed in Figure \ref{fig:population} as a signal on a graph. Observe that the estimates are reasonable since most of the resident population in London is concentrated toward the more remote areas of Zones 2--3, while business districts at the centre of Zone 1 are sparsely populated in the evening.


\vspace{-0.5cm}

\hspace{-0.8cm}
\begin{minipage}[t]{0.5\textwidth}
	\begin{figure}[H]
		\centering
		\includegraphics[width=1\textwidth, trim={0.3cm 0.4cm 0.3cm 0.4cm}, clip]{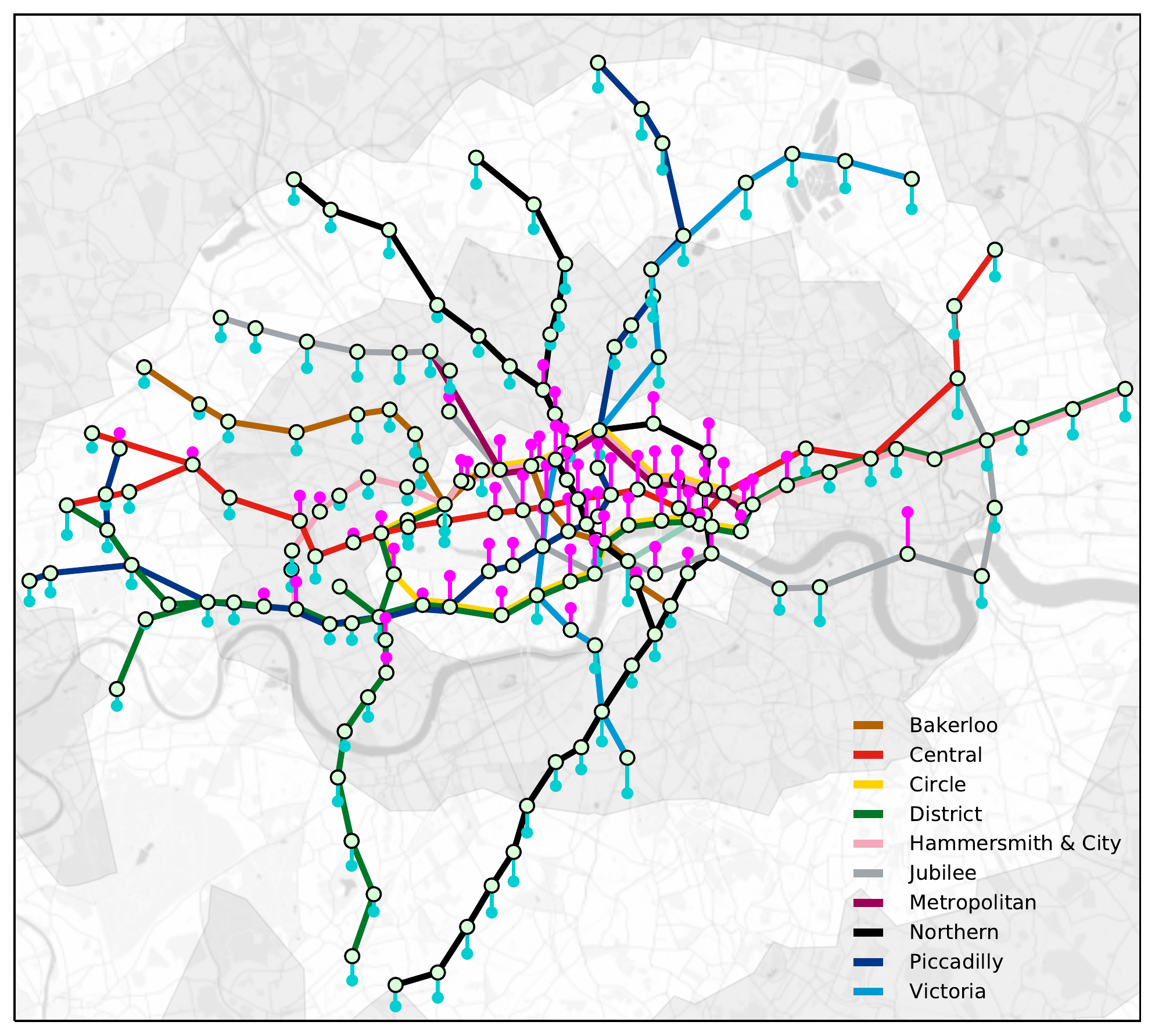}
		\caption{\label{fig:tube_net_flow}Net passenger outflow during morning rush hour within the Zones 1--3 of the London underground network. The magenta bars designate a net outflow of passengers while the cyan bars designate a net inflow of passengers. Stations located within business districts exhibit the greatest net outflow of passengers, while stations located in residential areas toward the fringes of Zones 2--3 exhibit the largest net inflow of passengers.}
	\end{figure}
\end{minipage}
\hspace{0.2cm}
\begin{minipage}[t]{0.5\textwidth}
	\begin{figure}[H]
		\centering
		\includegraphics[width=1\textwidth, trim={0.3cm 0.4cm 0.3cm 0.4cm}, clip]{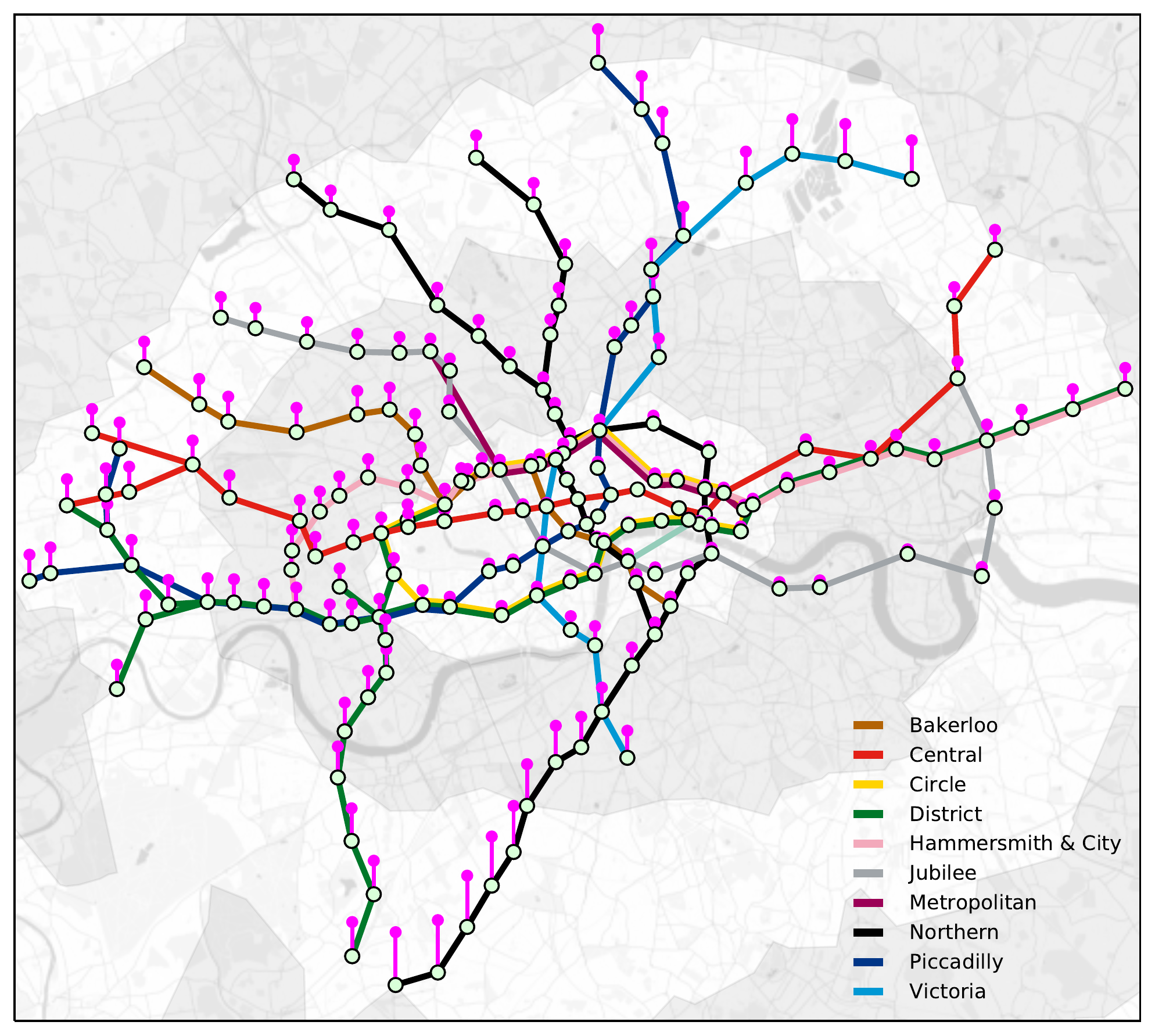}
		\caption{\label{fig:population}Population implied by our graph model in (\ref{eq:population_estimate}), from the net passenger outflow during morning rush hour within Zones 1--3. As expected, business districts exhibit the lowest population density, while residential areas (Zones 2--3) exhibit the highest commuter population density.}
	\end{figure}
\end{minipage}

\begin{minipage}[t]{0.45\textwidth}
	\begin{table}[H]
		\setlength{\tabcolsep}{7pt}
		\renewcommand{\arraystretch}{1.2}
		\begin{center}
			\caption{\label{table:netflows}Daily average passenger flows during the morning rush hour per Zone.}
			\begin{tabular}{c|rrr}
				\hline
				\textbf{Zone} & \textbf{Entries} & \textbf{Exits} & \textbf{Net Outflow} \\
				\hline
				$1$    & $455,704$ & $844,123$ & $388,419$  \\
				$2$    & $343,145$ & $264,732$ & $-78,413$  \\
				$3$    & $275,965$ & $104,414$ & $-171,551$ \\ \hline
				$4$--$10$    & $206,408$ & $72,152$  & $-134,256$  \\\hline
				\textbf{Total} & $1,281,222$ & $1,285,421$ & $4,199$ \\
				\hline
			\end{tabular}
		\end{center}
	\end{table}
\end{minipage}
\hspace{0.4cm}
\begin{minipage}[t]{0.45\textwidth}
	\begin{table}[H]
		\setlength{\tabcolsep}{7pt}
		\renewcommand{\arraystretch}{1.2}
		\begin{center}
			\caption{\label{table:netflows_top5}Stations with most net passenger outflow and inflow during the morning rush hour.}
			\begin{tabular}{l|r r r}
				\hline
				\textbf{Station} & \textbf{Entries} & \textbf{Exits} & \textbf{Net Outflow} \\
				\hline
				Bank          & $17,577$   & $69,972$  & $52,395$   \\
				Canary Wharf  & $8,850$    & $56,256$  & $47,406$   \\
				Oxford Circus & $3,005$    & $44,891$  & $41,886$   \\
				Green Park    & $2,370$    & $30,620$  & $28,250$   \\
				Holborn       & $1,599$    & $25,294$  & $23,695$   \\ \hline
				Finsbury Park & $20,773$   & $8,070$   & $- 12,703$ \\
				Canada Water  & $31,815$   & $14,862$  & $- 16,953$ \\
				Brixton       & $24,750$   & $4,369$   & $- 20,381$ \\
				Stratford     & $43,473$   & $22,360$  & $- 21,113$ \\
				Waterloo      & $61,129$   & $22,861$  & $- 38,268$ \\\hline
			\end{tabular}
		\end{center}
	\end{table}
\end{minipage}

\section*{Conclusions}

We have introduced a graph-theoretic framework which makes it possible to assess the functional design and operation of public transport networks in a mathematically tractable and intuitive manner. The formulation allows not only an assessment of current functionality but also offers an opportunity to facilitate enhancements, assess vulnerabilities notwithstanding the possibility of a dynamic management of the network. Our focus has been on establishing a link between the laws of diffusive fields and the parameters of urban underground traffic. In this way, we have been able to use the ability of graphs to process data on irregular domains to introduce an intuitive model for the population migration during the rush hour. It is our hope to have provided a platform to approach both traffic planning and urban transition issues from a rigorous engineering perspective using graph signal processing techniques.


\section*{Authors}

\noindent \textbf{Bruno Scalzo Dees} (bruno.scalzo-dees12@imperial.ac.uk) received the M.Eng. degree in aeronautical engineering from Imperial College London, United Kingdom. He is currently working toward the Ph.D. degree at the Department of Electrical Engineering at the same institution. His research interests include statistical signal processing, maximum entropy modelling and multilinear algebra.

\smallskip

\noindent \textbf{Anthony G. Constantinides} (a.constantinides@imperial.ac.uk) is an emeritus professor at Imperial College London, United Kingdom. He has been actively involved in research on various aspects of digital signal processing and digital communications for more than 50 years. He is a Fellow of the IEEE and the Royal Academy of Engineering and the 2012 recipient of the IEEE Leon K. Kirchmayer Graduate Teaching Award.

\smallskip

\noindent \textbf{Danilo P. Mandic} (d.mandic@imperial.ac.uk) is a professor of signal processing at Imperial College London, United Kingdom. He is a recipient of the 2019 Dennis Gabor award for outstanding achievements in neural engineering, and the 2018 Best Paper Award in IEEE Signal Processing Magazine. His contributions were recognized with the President's Award for Excellence in Postgraduate Supervision at Imperial College in 2014. He is a Fellow of the IEEE.


\footnotesize

\bibliographystyle{IEEEtran}
\bibliography{{./Bibliography}} 

\begin{thebibliography}{1}
\providecommand{\url}[1]{#1}
\csname url@samestyle\endcsname
\providecommand{\newblock}{\relax}
\providecommand{\bibinfo}[2]{#2}
\providecommand{\BIBentrySTDinterwordspacing}{\spaceskip=0pt\relax}
\providecommand{\BIBentryALTinterwordstretchfactor}{4}
\providecommand{\BIBentryALTinterwordspacing}{\spaceskip=\fontdimen2\font plus
\BIBentryALTinterwordstretchfactor\fontdimen3\font minus
  \fontdimen4\font\relax}
\providecommand{\BIBforeignlanguage}[2]{{%
\expandafter\ifx\csname l@#1\endcsname\relax
\typeout{** WARNING: IEEEtran.bst: No hyphenation pattern has been}%
\typeout{** loaded for the language `#1'. Using the pattern for}%
\typeout{** the default language instead.}%
\else
\language=\csname l@#1\endcsname
\fi
#2}}
\providecommand{\BIBdecl}{\relax}
\BIBdecl

\bibitem{Freeman1977}
L.~C. Freeman, ``A set of measures of centrality based on betweenness,''
  \emph{Sociometry}, vol.~40, pp. 35--41, 1977.

\bibitem{Brandes2005}
U.~Brandes, \emph{Network Analysis: Methodological Foundations}.\hskip 1em plus
  0.5em minus 0.4em\relax Springer, 2005.

\bibitem{TFL}
\BIBentryALTinterwordspacing
Transport for london. [Online]. Available: \url{https://tfl.gov.uk/}
\BIBentrySTDinterwordspacing

\end{thebibliography}

\end{document}